\documentclass[twocolumn,showpacs,preprintnumbers,amsmath,amssymb]{revtex4}

\usepackage{graphicx}
\usepackage{epstopdf}
\usepackage{color}

\begin{document}


\title{Reproducible domain wall pinning by linear non-topographic features in a ferromagnetic nanowire}

\author{M. A. Basith}
\email[Author to whom correspondence should be addressed (e-mail): ]{m.basith@physics.gla.ac.uk}
\affiliation{
School of Physics \& Astronomy, University of Glasgow, G12 8QQ, United Kingdom.
}
\author{S. McVitie}
\affiliation{
School of Physics \& Astronomy, University of Glasgow, G12 8QQ, United Kingdom.
}
\author{D. McGrouther}
\affiliation{
School of Physics \& Astronomy, University of Glasgow, G12 8QQ, United Kingdom.
}
\author{J. N. Chapman}
\affiliation{
School of Physics \& Astronomy, University of Glasgow, G12 8QQ, United Kingdom.
}

\begin{abstract}
We demonstrate that for multilayered magnetic nanowires, where the thickness and composition of the individual layers has been carefully chosen, domain walls can be pinned at non-topographic sites created purely by ion irradiation in a focused ion beam system. The pinning results from irradiation induced alloying leading to magnetic property modification only in the affected regions. Using Lorentz transmission electron microscopy, we have studied the pinning behavior of domain walls at the irradiation sites. Depending on the irradiation dose, a single line feature not only pinned the domain walls but also acted to control their structure and the strength of their pinning.\\

\end{abstract}

\maketitle	
The ability to control precisely the motion of magnetic domain walls (DWs) in ferromagnetic nanowires is crucial for the future realization of proposed magneto-electronic (or spintronic) devices \cite{ref5.1, ref5.2} as well as for greater understanding of fundamental nanomagnetic behavior \cite{ref5.3}. Previous investigations have demonstrated that topographically defined trapping sites such as notch/anti-notch of different geometries allow control of the position of DWs \cite{ref5.7, ref5.77, ref5.5, ref5.6, ref5.32,ref5.33} in ferromagnetic nanowires. The strength of these trapping/pinning features depends on the feature geometry dimensions and the spin distribution of the incoming DWs \cite{ref5.5, ref5.6,ref5.33}. A recent investigation revealed  that the depinning field of DWs in Permalloy (Py) nanowires can be controlled by interactions between magnetic stray fields and artificial constrictions \cite{ref5.76}. However, as the width of the nanowires and the geometrical pinning features is reduced, the structure of the wire edges as well as the features edges becomes increasingly important \cite{ref5.28, ref5.29}. Width reduction, particularly below 100 $nm$, and control of well defined edge structure present major challenges for current fabrication techniques. A possible alternative approach is to use ion irradiated features \cite{ref5.30, ref5.31} that controllably modify the properties of the film in selected regions within the patterned structures. The modified areas may act as inherent pinning sites without necessarily changing the geometry of the nanowires. Notably, focused ion beam (FIB) implantation/irradiation has been shown to be capable of modifying magnetic properties such as the coercivity, saturation magnetization M${_s}$ \cite{ref5.17,ref5.18} and local anisotropy direction \cite{ref5.19} in continuous Py films. In multilayered ferromagnetic films, where interfaces are present, magnetic properties have been shown to be even more sensitive to FIB irradiation \cite{ref5.38,ref5.39,ref5.40}. Moreover, in a recent investigation \cite{ref5.41}, tailoring of the magnetic properties of layered magnetic films Ni$_{81}$Fe$_{19}$/Au by radiation-induced modification of interfaces was also explored. That study demonstrated the ability of a FIB to modify locally the magnetic properties of ultra thin in-plane magnetized capped films due to ion beam mediated intermixing of the layers.

\begin{figure}[hh]
\centering
\includegraphics[width=5cm]{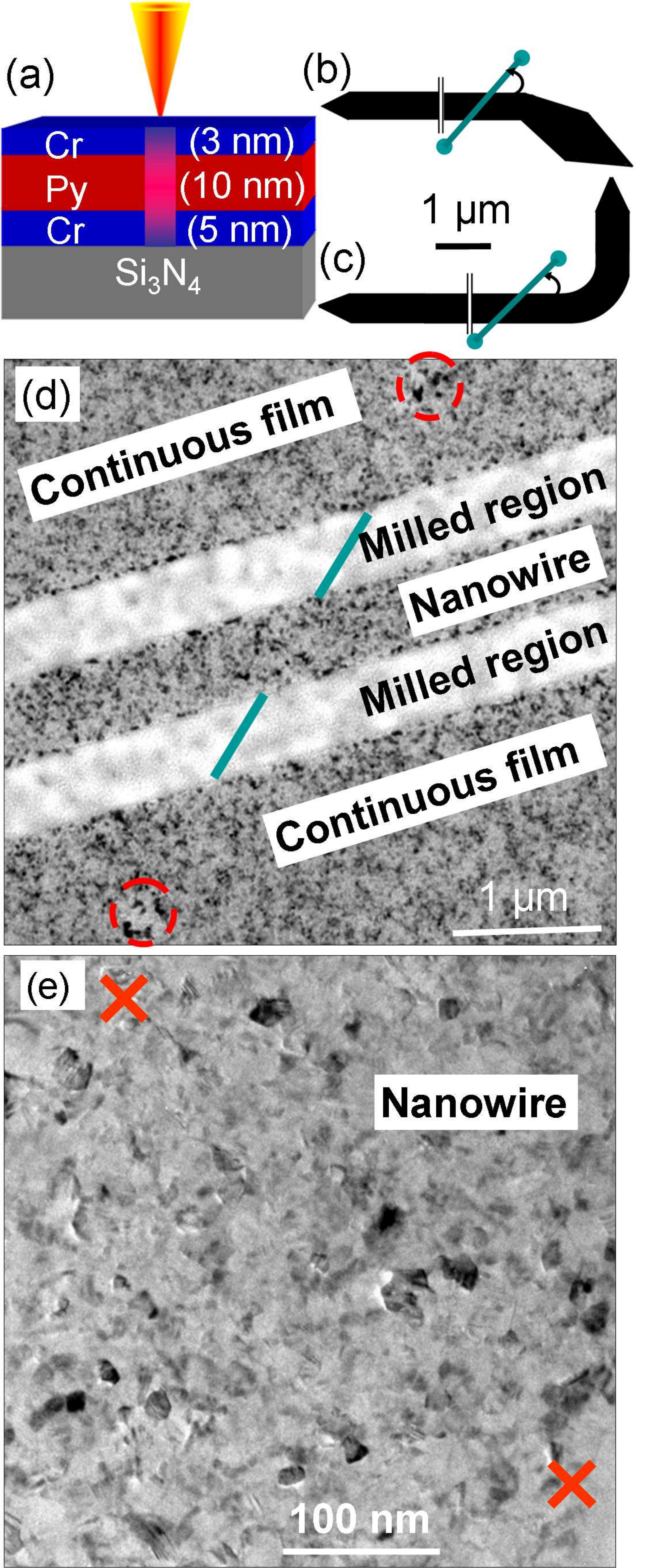}
 \caption { (Color online) (a) Schematic showing the cross-section of the multilayer thin film stack indicating local alloying due to the FIB irradiation. Note that the color variation is intended as a schematic representation only. (b and c) are plan view schematics of the nanowire geometries used which show the location of the FIB irradiated line. The width of the nanowires is 500 $nm$. The length of the nanowires has been shortened schematically as indicated by the vertical bars. (d) Plan view TEM image shows the nanostructures written by FIB milling on multilayer thin film stack consists of Cr(3 $nm$)/Py(10 $nm$)/Cr(5 $nm$). The nanowire has been written from a continuous film stack and the regions of nanowire and remaining continuous film are shown. The cyan line shows the orientation of the irradiated line written continuously between the two red circles. (e) Plan view TEM image at the centre of the nanowire containing the irradiated line. The location of the line is between the two "x" points on the image.} \label{fig1}  
\end{figure}

In this letter, we demonstrate the novel creation of non-topographic DW pinning sites, defined directly by FIB irradiation, where the pinning strength and structure of the DWs pinned is strongly controlled by the irradiation dose. In the present investigation, a tri-layer thin film of Cr/Py/Cr was purposely created to optimize sensitivity to ion irradiation through localised beam induced alloying of the Py and Cr in a FIB microscope. The structure of the multilayer system consists of a magnetic thin film layer of Py surrounded by nonmagnetic layers on top and bottom of the magnetic layer. The nonmagnetic metallic layers could be any metal e.g. Al, Au, Cu, Cr. In this study Cr was chosen because alloying Cr with Py is known to significantly reduce the Curie temperature, M${_s}$, magnetic anisotropy as well as altering the exchange constant \cite{ref5.22,ref5.21,ref5.20} of Py for only a small addition of Cr. A mixing of around 8 at\% Cr is sufficient to render the Py/Cr alloy paramagnetic at room temperature \cite{ref5.24}. The thickness of each layer of the tri-layer system was optimized by studying results from the dynamic ion irradiation simulation package TRIDYN \cite{ref5.25}. These simulations showed that an asymmetric thickness of the top and bottom Cr layer yielded the most uniform Cr concentration, resulting directly from the distribution of energy lost with depth of the implanted ions. Experiments to investigate local alloying due to interfacial atomic mixing were performed using FIB irradiation as shown schematically in Fig. \ref{fig1}(a). The pinning/de-pinning of DWs in the patterned nanowires at/from ion irradiated lines of varying doses was investigated using the Fresnel mode of Lorentz transmission electron microscopy (LTEM) \cite{ref5.26}. \emph{In situ} LTEM imaging was performed in a Philips CM20 field emission gun TEM equipped with Lorentz lenses and customised for performing \emph{in situ} magnetizing experiments \cite{ref5.57}. 

A tri-layer thin film of Cr(3 $nm$)/Py(10 $nm$)/Cr(5 $nm$) was deposited by magnetron sputtering without breaking vacuum at a rate not exceeding 0.55 $\ \AA$/s  on top of Si$_{3}$N$_{4}$ window membrane suitable for TEM experiments \cite{ref5.42}. During the deposition of the magnetic material an in-plane field was applied to produce a uniaxial magnetic anisotropy along the length of the wires. Magnetooptic Kerr effect magnetometry showed a well defined uniaxial anisotropy of the as-deposited film. The anisotropy field H$_K$ and the coercivity H$_C$ of the continuous film were measured to be $3.9 $ Oe and $2.2$ Oe, respectively. Thereafter, isolated nanowires  with the long axis of the nanowires aligned parallel to the anisotropy axis of the Py film were patterned on the continuous film by FIB etching (using ion doses in excess of $10$x$10^{15} \ ions/cm^2$ where physical sputtering of the film occurred). The thickness of the planar Py nanowire with a rectangular cross-section is 10 $nm$, the width is 500 $nm$ and the length is 15 $\ \mu m$. The structures of the fabricated 500 $nm$ wide nanowires are shown schematically in Figs. \ref{fig1}(b) and (c). The right-most end of the nanowire was connected to a diamond shaped pad Fig. \ref{fig1}(b) or a right-angled bend Fig. \ref{fig1}(c) \cite{ref5.5} to allow control over the formation of DWs with an external field. The left-most nanowire end was tapered to prevent nucleation of DWs from this end.  In each nanowire, the irradiated line was written at an angle of $45^{\circ}$ from the wire axis, as shown in the schematic Figs. \ref{fig1}(b) and (c). At the ends of the cyan lines, Figs. \ref{fig1}(b) and (c), larger spots were positioned to provide visible markers identifying the line's location during TEM investigation. Patterning of the isolated nanowires and irradiation of pinning sites and marker spots were performed using a 10 pA aperture of an FEI Nova NanoLab 200 SEM/FIB workstation (Ga$^+$ source operated at 30 keV). The linear pinning sites were created by scanning the ion beam once only along a line and, by varying the beam dwell time, irradiation doses could be controlled for these lines. Nominally the doses of the irradiated lines corresponded to $d$x$10^{15} \ ions/cm^2$ (with $d$ = 4, 8, 12, 16 and 20). Three wires of each structure were written using exactly the same ion beam conditions and irradiation line doses. Figure \ref{fig1}(d) shows a plan view TEM bright field (BF) image of a patterned nanowire in which the pinning site was irradiated with the highest dose of $20$x$10^{15} \ ions/cm^2$. Heavily irradiated marker spots of diameter 300 $nm$ were indicated by red circles in Fig. \ref{fig1}(d). The nanowire has been written from the continuous film stack and the regions of nanowire and remaining continuous film are shown in Fig. \ref{fig1}(d). A higher magnification BF image, Fig. \ref{fig1}(e), was recorded from the centre of the nanowire along which the irradiation line was written and the location of the line is indicated between the two "x" points on the image. It is apparent, and a little surprising, that even at this high dose level no visible effect on the grain structure of the nanowire could be discerned in high magnification TEM images.

\begin{figure}[!hh]
\centering
\includegraphics[width=6.5cm]{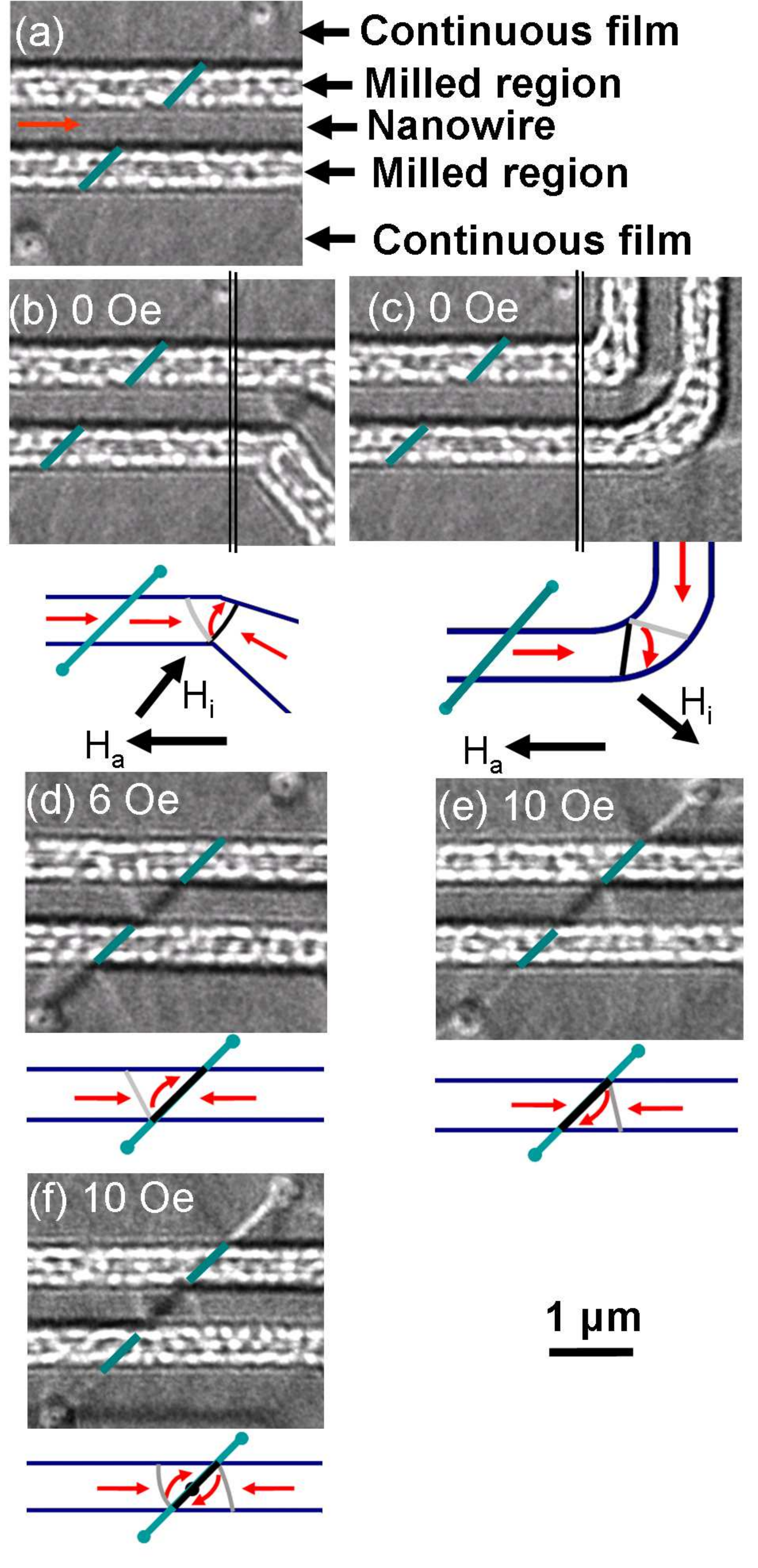}
\caption{ (Color online) (a) Fresnel image showing a uniformly magnetized nanowire with magnetization direction pointing from left to right as marked by red arrow. Fresnel images (b and c) show asymmetric TDWs in the nanowires with magnetization pointing up and down, respectively at the center of the wall at zero field. The length of the patterned nanowire has been reduced as indicated by vertical bars. (d and e) show pinning of these asymmetric TDWs at the FIB irradiated sites for line doses of $4$x$10^{15} \ ions/cm^2$ and $8$x$10^{15} \ ions/cm^2$,  respectively. (f) Fresnel image shows the transformation of an initial asymmetric TDW into a clockwise (cw) VDW at the pinning site where the dose was $16$x$10^{15} \ ions/cm^2$. The schematic diagrams show the magnetization distribution deduced from the corresponding images. Schematic diagrams associated with images (b) and (c) also indicate the direction of the initial field, H$_{i}$, used to create the DWs and the applied field, H$_{a}$, to propagate the DW's, respectively in each case.} \label{fig2} 
\end{figure}

In the next stage of this investigation, \emph{in situ} magnetization reversal experiments were performed using the Fresnel mode of Lorentz TEM to observe the reproducibility and controlled pinning and depinning of the DWs along the patterned nanowires. The two different right-most wire end structures, schematics Figs. \ref{fig1}(b) and (c), were employed so that during TEM investigations, DWs could be nucleated with the same head-to-head magnetization sense but with opposite chirality. The Fresnel image, Fig. \ref{fig2}(a), shows a uniformly magnetized nanowire with magnetization pointing to the right as marked by red arrow. The nanowire, milled regions and remaining continuous film are clearly indicated in Fig. \ref{fig2}(a). Figures \ref{fig2}(b and c) show that field induced nucleation resulted in DWs of asymmetric transverse structure (TDW) and with magnetization pointing up/down at the centre of the wall, controlled by the diamond shaped pad/right-angled bend end geometry. Schematic diagrams show the magnetization distribution deduced from each of the Fresnel images. Formation of DWs in the wire structure, Fig. \ref{fig1}(b) was achieved by applying an initial magnetic field (H$_{i}$, schematic Fig. \ref{fig2}(b)) at an angle offset from  perpendicular to the wire axis and then relaxing the field to zero \cite{ref5.27}. Whereas in wire structure, Fig. \ref{fig1}(c), DWs were formed by applying an initial magnetic field H$_{i}$ as shown in the schematic Fig. \ref{fig2}(c) and relaxing the field to zero. For straight nanowires of 500 $nm$ width and 10 $nm$ Py thickness, the lowest energy wall structure, predicted from the DW phase diagram \cite{ref5.34} and our own micromagnetic simulations performed using object oriented micromagnetic framework (OOMMF) \cite{ref5.36}, should be a vortex structured DW (VDW). However, the nature of the field application to form the DWs meant that asymmetric TDWs were always observed after the nucleation step. According to simulations performed on the OOMMF package the energy difference between asymmetric TDWs and VDWs for walls in straight nanowires with the dimensions considered here is only a few percent. Given that the energy difference between the two DW types is small in a straight nanowire and considering the walls are formed in a region of the nanowire with varying or non-straight edges and a strong polarizing field, it is perhaps not surprising that the asymmetric TDWs are present initially in the nanowires. 

\begin{table}[!h]
\caption{The table shows the wall type on arriving at the pinning site irradiated as a function of different ion doses. DW depinning fields as a function of irradiation doses for the different types of initial wall structures are also shown.}  
\begin{center}
\begin{tabular}{|l|l|l|l|l|}
 \hline
{Dose : } &\multicolumn{2}{c|}{Initial asymmetric TDW}& \multicolumn{2}{c|}{Initial asymmetric TDW} \\
{$d\times 10^{15}$}  &\multicolumn{2}{c|}{structure: magnetization}& \multicolumn{2}{c|}{structure: magnetization} \\
{$\ ions/cm^2$}  &\multicolumn{2}{c|}{points up (\includegraphics[width=0.25cm]{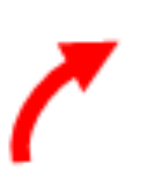})}& \multicolumn{2}{c|}{points down (\includegraphics[width=0.25cm]{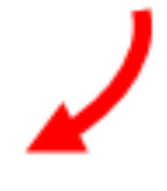})} \\
\cline{2-5}
 &DW pinned at&Depin. &DW pinned at &Depin.\\
 &pinning site as a:&field &pinning site  as a:&field\\ 
 &&(Oe)&&(Oe)\\

\hline
4&TDW: \includegraphics[width=0.25cm]{Fig3.eps} (40\%)&$11\pm1$&TDW: \includegraphics[width=0.25cm]{Fig3.eps} (40\%)&$16\pm2$\\
&TDW: \includegraphics[width=0.25cm]{Fig4.eps} (60\%)&$16\pm4$&TDW: \includegraphics[width=0.25cm]{Fig4.eps} (60\%)&$20\pm4$\\
\hline
8&TDW: \includegraphics[width=0.25cm]{Fig3.eps} (20\%)&$65\pm8$&TDW: \includegraphics[width=0.25cm]{Fig3.eps} (25\%)&$55\pm4$\\
&TDW: \includegraphics[width=0.25cm]{Fig4.eps} (80\%)&$79\pm10$&TDW: \includegraphics[width=0.25cm]{Fig4.eps} (75\%)&$64\pm8$\\
\hline
12&TDW: \includegraphics[width=0.25cm]{Fig4.eps} (14\%)&$76\pm4$&TDW: \includegraphics[width=0.25cm]{Fig4.eps} (80\%)&$80\pm4$\\
&VDW: cw (86\%)&$75\pm2$&VDW: cw (20\%)&$76\pm4$\\
\hline
16&VDW: cw (100\%)&$79\pm2$
&VDW: cw (100\%)&$79\pm4$\\
\hline
20&VDW: cw (100\%)&$82\pm4$
&VDW: cw (100\%)&$82\pm4$\\
\hline
\end{tabular}
\end{center}
\end{table}
After nucleation, DWs were moved by applying a magnetic field (H$_{a}$, schematics associated with Figs. \ref{fig2}(b) and (c)) opposite to the direction of the magnetization in the nanowire. The initial asymmetric TDWs moved from the corner and were subsequently pinned at the irradiated lines as shown in the Fresnel images and their corresponding schematics in Figs. \ref{fig2}(d and e), respectively. The experiments carried out throughout this investigation were repeated ten times each for six wires of each irradiation line dose. Each set of measurements commenced from the states shown in Figs. \ref{fig2}(b and c) and then followed the sequence propagation, pinning and depinning with increasing applied field. The initial states of the wall structures, Figs. \ref{fig2}(b and c) were then reset and the measurements were repeated. The strength of the magnetic field was increased gradually and the movement of the DW from the corner position was noted. The propagation field i.e. the field required to move the DW from the corner position to the pinning site was $8\pm2$ Oe and the wall structure was observed to be stable up to this point. On reaching the pinning site the wall structure was observed and it was noted that at lower doses the wall appeared to remain transverse in character as seen in  Figs. \ref{fig2}(d and e). However there was a tendency for the transverse wall to have a downwards pointing domain, Fig. \ref{fig2}(e), irrespective of the initial wall state. This tendency became more pronounced at higher doses. Furthermore at the highest doses only a clockwise vortex domain wall was observed once the domain wall was pinned at the irradiated site, as seen in the Fresnel image and its corresponding schematic, Fig. \ref{fig2}(f). Table 1 summarizes the results for each initial wall type and the subsequent pinned wall structure and its depinning field as a function of the line dose. Although quite a range of behavior was observed some clear patterns of dependency can be deduced. Irrespective of the direction of the central magnetization (up/down) of the initial wall at nucleation, at the pinning site the asymmetric TDW with downwards pointing magnetization was most often observed up to a dose of $8$x$10^{15} \ ions/cm^2$. Furthermore, the de-pinning field of this wall was seen always to be higher than that of the asymmetric TDW with the upwards magnetized central domain. These observations, and the measured de-pinning field strengths, show that the lines irradiated with the two lowest doses act as effective pinning sites for asymmetric TDWs of both upwards and downwards central magnetization but that the latter are pinned more strongly. Owing to the quasi-static manner of applying the propagating field and its low value, we believe that the asymmetric TDW up/down transformations are mediated by the pinning site itself rather than occurring during propagation. At higher doses, $12$x$10^{15} \ ions/cm^2$ and greater, the lower energy VDWs became the dominant structure observed at the pinning site. 

Additionally the depinning field for the VDW appeared to be fairly independent of the ion dose of the pinning site. Clearly the irradiated line (width in the range 30-60 $nm$), which is very much narrower  than the effective width of either of the asymmetric TDW or VDW types, exerts a significant influence over which type is most energetically favorable, for irradiation doses $>$ $12$x$10^{15} \ ions/cm^2$. As stated previously, the effect of the irradiation results in local alloying \cite{ref5.38,ref5.39,ref5.40,ref5.41,ref5.22,ref5.21,ref5.20} and modifications in the magnetic moment strength, the anisotropy, the exchange constant, and the Curie temperature along the irradiated line. This provides a possible mechanism then for the asymmetric TDW, when arriving at the pinning site, to convert to a lower energy VDW type. As mentioned earlier, micromagnetic simulations are highly instructive showing that for a free nanowire, both asymmetric TDWs and VDWs have energies within a few percent of each other but that 80-90 \% of the total energy cost of each type results from demagnetizing fields. However, the distributions of the sources of demagnetizing field are quite different for both asymmetric TDW and VDW types. For the asymmetric TDW significant demagnetizing fields are generated by the two low angle walls enclosing the central upwards/downwards domain and also at their intersections with the wire edge. For the VDW, the strong sources of demagnetizing field are the vortex core itself and also the ends of $180^{\circ}$  N\'eel Wall. These features can all lie along the irradiated line and demagnetizing energies scale with M${_s^2}$, our OOMMF simulations show that the demagnetizing energy of the VDW is reduced when it is located at the irradiated line.


In conclusion, we observe that the inclusion of the non topographic pinning features along the length of the patterned nanowires allow a degree of control of pinning and depinning of DWs at predefined locations. These pinning features behave like potential wells and the strength of the wells is seen to increase up to the formation of the VDWs at which point it appears to plateau. Wall transformations are observed with a tendency of one type of asymmetric TDW wall favored at lower doses and only VDWs at the highest doses. In this sense the pinning site can be seen as setting the DW type for the higher doses whilst increasing the probability of one type of asymmetric TDW at the lower doses. Micromagnetic simulations performed using OOMMF package indicate that the irradiated lines reduce the demagnetizing field in the DWs when they are located at these positions. In addition to the control of DWs, our findings highlight the potential for engineering or filtering DWs of certain structure types as they pass linear features irradiated at specific doses at chosen locations along patterned nanowires. What we have demonstrated here is the ability to pin a domain wall at a linear feature rather than a block region \cite{ref5.30, ref5.31}. The wall type at the pinning feature appears to be controlled by the dose and the strength of pinning has also been measured in the system we have chosen. We are carrying out further work to further investigate the dependence of this pinning and its effect on the wall structure at the defect in terms of the defect geometry.

The authors would like to acknowledge R. Mattheis from IPTH Jena, Germany for film deposition and T. Strache from HZDR, Dresden, Germany for MOKE investigation of the as-deposited film. 


\end{document}